# Design of Silicon Slab Waveguides based all-optical Logic Gates

**Anmol Aggarwal[a], Ashi Mittal[a], and Yogita Kalra[a],***

[a] *Advanced Simulation Lab, TIFAC-Centre of Relevance and Excellence in Fiber Optics and Optical Communication, Department of Applied Physics, Delhi Technological University (Formerly) Delhi College of Engineering, Bawana Road, Delhi 110042, India.*
* Corresponding author (Email: dryogitakalra@gmail.com; dryogitakalra@dtu.ac.in)

**Short Title: SSW-based all-optical Logic Gates**

**Abstract:** Logic gates are an indispensable part of digital circuits, and therefore such devices with fast operating speeds and simple designs are vital for all kinds of electronic applications. This paper proposes simple Y-shaped designs of all-optical AND, OR, and NOT logic gates in two dimensions that are operational in both TE and TM modes. These gates are designed using silicon slab waveguides, which makes them relatively easy to fabricate. The logic gates are designed, analyzed, and optimized at 1.55 µm using the finite difference time domain method. The truth table of the proposed all-optical logic gates is realized for every input case. The propagation delay time is determined to be of the order of femtoseconds ($10^{-15}$ s), which is very fast compared to the response time of the currently existing all-optical logic gates ($10^{-9}$ s - $10^{-12}$ s).

**Keywords: Photonics, Logic Gates, Silicon Slab Waveguides, Optical Communication, Optical Circuits.**

## Acknowledgments

The authors gratefully acknowledge the initiatives and support toward the establishment of "TIFAC CENTRE of Relevance and Excellence in Fiber Optics and Optical Communication at Delhi Technological University, formerly Delhi College of Engineering" through the "Mission REACH" program of Technology vision-2020, Government of India.

# Introduction

Logic gates are the devices that are used for making logical decisions by a computer/microprocessor [1, 2]. To enhance the performance of such circuits, all-optical logic gates can be employed to substantially increase their operational speed.

The currently devised designs use several technologies, which include, semiconductor optical amplifier (SOA) [3-5], highly nonlinear fiber (HNLF) [6-8], interferometric structures [9, 10], periodically poled lithium niobate (PPLN) waveguides [11-13], plasmonic waveguide-based designs [14-18], topological photonics-based designs [19-21], artificial neural networks based designs [22-24], etc. All these technologies have their merits but most of them suffer from excess power consumption and bulky designs. PPLN, topological photonics, and plasmonic-based waveguides also face issues like dependencies on temperature, phase, polarization, and high loss of signal, which reduces their performance [25].

Kita et al. proposed a logic gate design based on silicon wires [26]. The proposed design uses three input ports and a single design for many binary logic operations based on the variation of phase and power of the input sources. Another widely investigated method to realise all-optical logic gates is the use of photonic crystals (PhC) for guiding the light as done by [27-36].

In this work, the proposed all-optical logic gates use silicon slab waveguides (SSWs). The light is guided in SSWs because of the total internal reflection (TIR) of light into the waveguide. The higher the difference between the refractive index of the material and the medium outside it, the lower the critical angle. Whenever the angle of incidence of light on the waveguide-air interface is greater than the critical angle, the light is reflected internally and hence guided by the waveguide. The refractive index of silicon is quite high at 1.55 µm which ensures a low critical angle and low leakage.

One of the main motivations behind experimenting with silicon as a choice of material for our slab waveguides is that the fabrication process of optoelectronic devices is widely explored with its peculiarities known. Hence, considering the future application of such devices, silicon, in our opinion could be ideal [37, 38]. The SSWs in the proposed structures are surrounded by air and have strong optical confinement. They work in both transverse electric (TE) and transverse magnetic (TM) modes, which is not the case with Kita et al's design [26]. The gates have been optimized for a light of wavelength 1.55 µm at which the refractive index of silicon is 3.5, with the help of the finite difference time domain method (FDTD) software MEEP [39]. The FDTD method relies on Maxwell's equations to find a solution, the method is scalable and does not involve any physical approximations. These are some of the many advantages due to which this method has been employed [39].

The design of the gates has been set such that the single-mode condition is satisfied. They are easy to fabricate because of their simple design. The propagation losses for SSWs have been determined to be 0.1 dB/mm for TE mode [40, 41] and 0.09 dB/mm for TM mode [41]. These minute loss values further prompted us to work with SSWs. To test their performance, we have realized the truth table by computing the normalized output power and determined the propagation time delay and decibel losses for each gate. Further, due to their simple structure, the proposed all-optical logic gates are faster as compared to the current logic gates, including PhC-based designs, which report a response time in the range of a few nanoseconds or pico-seconds [27-30, 33, 34, 42, 43].

# Design of the Y-shaped logic gates

Three SSWs have been used to create a Y-shaped structure to realize the logic gates. This specific design was selected as our gates are binary logic gates and require two input ports and one output port. Further, Y-shaped waveguides are widely used in all-optical photonic crystal logic gates [33, 34].

An air hole has been introduced at the intersection of the waveguides. The objective of the air hole is to create an appropriate phase variation to enhance and get an optimised output. **Fig. 1(a)** depicts the schematic structure of the proposed logic gate design, with two input and one output port.

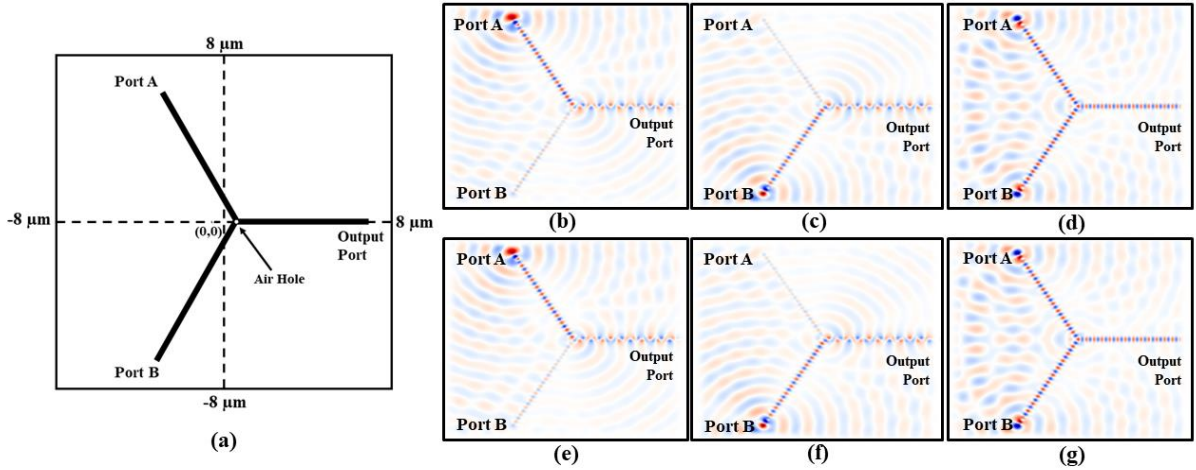

**Fig. 1:** (a) The schematic structure of the basic logic gates created using SSWs. The intensity distributions of the AND logic gate in the TE mode of propagation (b) A=1, B=0, (c) A=0, B=1, (d) A=1, B=1; and the TM mode of propagation (e) A=1, B=0, (f) A=0, B=1, (g) A=1, B=1

The sources used at ports A and B are continuous and point size. A 2-dimensional (2D) 16 µm ×16 µm structure has been simulated in the x-y plane. The length of the SSWs has been set to be 7 µm. They are inclined at an angle of 120° with respect to each other. The perfectly matched layers (PML) act as the boundary layers of the structure. The width of the SSWs has been set to 0.241 µm for TE mode propagation and 0.345 µm for TM mode propagation to satisfy the single mode condition, which is given by **Eqn. 1**.

$$\nu_c = \frac{c}{2d\sqrt{n_1^2 - n_2^2}} \quad (1)$$

Here 'd' is the width of the SSW, $n_1$ is the refractive index of silicon at frequency ν, where ν is the frequency of the input light; $n_2$ is the refractive index of air, and c is the speed of light in vacuum. To satisfy the single-mode condition, ν should be greater than $\nu_c$.

The width of the waveguide for the TE mode has been optimised at 0.241 µm which is less than the optimised width for the TM mode. This can be attributed to the fact that when the TE mode is used, the effective refractive index (ERI) at the interface is higher as compared to the one when TM mode is used. This provides stronger optical confinement and hence a smaller width.

The ratio of the output power ($P_o$) to the input power ($P_{in}$) has been determined for each gate. If the ratio is greater than 0.5 then the output has been considered as logic one (1) else, it has been considered a logic zero (0). Most of the outputs are observed to be significantly higher than 0.5 for a true case and appreciably lower than 0.5 for a false case, which makes their detection and distinction easy.

## Design and Analysis of AND Gate

The AND gate has been realized in both TE and TM modes. For the TE mode, the radius of the hole has been optimized as 0.21 µm, which is positioned at 0.3 µm to the right of the origin. The radius of the hole has been taken as 0.18 µm for the TM mode, positioned at 0.4 µm to the right of the origin. **Table 1** depicts the truth table and losses of the logic gate in the TE and TM modes respectively which are calculated using **Eqn. 2**.

$$Loss = 10\, log_{10}\left(\frac{P_{out}}{P_{in}}\right) \qquad (2)$$

**Table 1:** The truth table and losses for AND logic gate in the TE and TM mode of propagation.

| Logical Operations | | | TE Mode | | TM Mode | |
|---|---|---|---|---|---|---|
| Logical Input at Port A | Logical Input at Port B | Logical Output | Normalized Output Power | Loss (dB) | Normalized Output Power | Loss (dB) |
| 0 | 0 | 0 | 0 | 0 | 0 | 0 |
| 0 | 1 | 0 | 0.3046 $P_{in}$ | 5.1627 | 0.3033 $P_{in}$ | 5.1813 |
| 1 | 0 | 0 | 0.3046 $P_{in}$ | 5.1627 | 0.3033 $P_{in}$ | 5.1813 |
| 1 | 1 | 1 | 0.6307 $P_{in}$ | 2.0018 | 0.7373 $P_{in}$ | 1.3236 |

The intensity distribution of the AND gate for different inputs for both TE and TM modes is presented in **Figs. 1(b), 1(c), 1(d), 1(e), 1(f), and 1(g)**. The truth table shows the realization of the AND gate. Further, the propagation delay time for both modes has been obtained.

The propagation time delay is the time taken by the structure to produce a stable output when the input is stable [44]. The propagation time delay for the TE mode of the AND gate has been determined as 8.17 femtoseconds and for the TM mode as 7.34 femtoseconds.

## Design and Analysis of OR Gate

The OR gate has also been realized in both TE and TM modes. For the TE mode, the radius of the hole has been optimized as 0.15 µm, which is positioned at 0.4 µm to the right of the origin. The radius of the hole is taken to be 0.21 µm for the TM mode. It is situated at 0.4 µm to the right of the origin. **Table 2** depicts the truth table and losses of this logic gate functioning in the TE and TM modes respectively which are calculated using **Eqn. 2**.

**Table 2:** The truth table and losses for OR logic gate in the TE mode of propagation.

| Logical Operations | | | TE Mode | | TM Mode | |
|---|---|---|---|---|---|---|
| Logical Input at Port A | Logical Input at Port B | Logical Output | Normalized Output Power | Loss (dB) | Normalized Output Power | Loss (dB) |
| 0 | 0 | 0 | 0 | 0 | 0 | 0 |
| 0 | 1 | 1 | 0.5658 $P_{in}$ | 2.4734 | 0.6587 $P_{in}$ | 1.8131 |

| 1 | 0 | 1 | 0.5658 $P_{in}$ | 2.4734 | 0.6587 $P_{in}$ | 1.8131 |
| 1 | 1 | 1 | 0.6851 $P_{in}$ | 1.6425 | 0.8189 $P_{in}$ | 0.8677 |

The intensity distribution of the OR gate for different inputs for both TE and TM modes has been presented in **Fig. 2**. The truth table shows the realization of the OR gate.

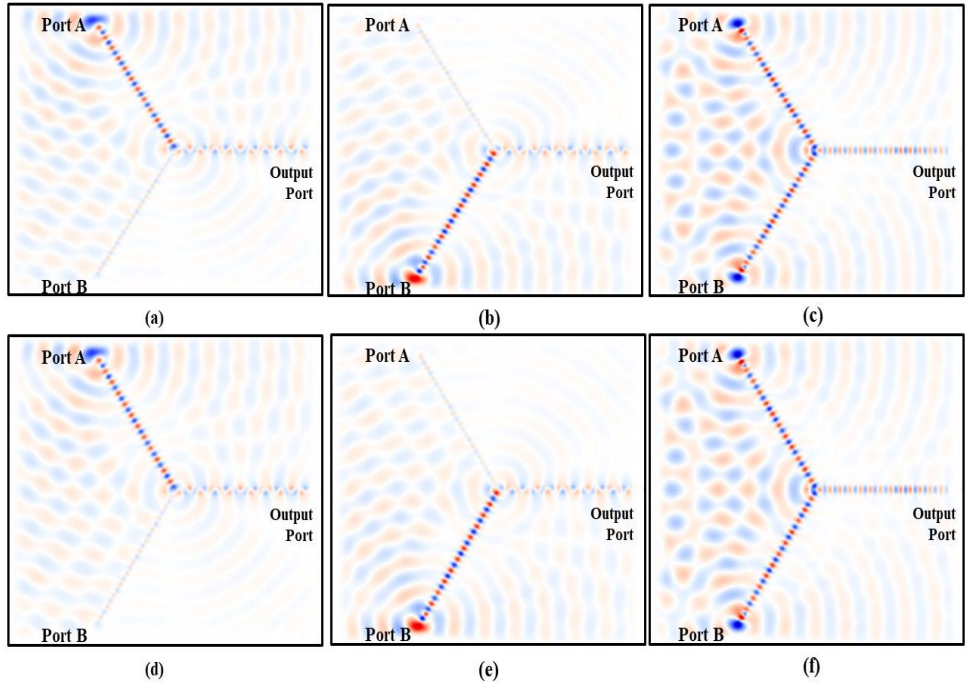

**Fig. 2:** The intensity distributions of the OR logic gate in the TE mode of propagation (a) A=1, B=0, (b) A=0, B=1, (c) A=1, B=1; and the TM mode of propagation (d) A=1, B=0, (e) A=0, B=1, (f) A=1, B=1

The propagation time delay for the TE mode of the OR gate has been determined as 8.67 femtoseconds and for the TM mode as 7.17 femtoseconds.

## Design and Analysis of NOT Gate

The NOT gate has also been realized in both TE and TM modes. For the TE mode, the radius of the hole has been optimized as 0.165 µm, which is positioned at 0.4 µm to the right of the origin and 0.07 µm above the x-axis. The radius of the hole has been taken as 0.245 µm for the TM mode. It is situated at 0.4 µm to the right of the origin and 0.02 µm below the x-axis. **Table 3** depicts the truth table and losses of this logic gate functioning in the TE and TM modes respectively which are calculated using **Eqn. 2**.

**Table 3:** The truth table and losses for NOT logic gate in the TE mode of propagation.

| Logical Operations | | | TE Mode | | TM Mode | |
|---|---|---|---|---|---|---|
| Logical Input at Port A | Logical Input at Port B | Logical Output | Normalized Output Power | Loss (dB) | Normalized Output Power | Loss (dB) |
| 1 | 0 | 1 | 0.8034 $P_{in}$ | 0.9507 | 0.6567 $P_{in}$ | 1.8263 |
| 1 | 1 | 0 | 0.4885 $P_{in}$ | 3.1114 | 0.4905 $P_{in}$ | 3.0936 |

The intensity distribution of the NOT gate is presented in **Fig. 3**. Port A acts as the reference port and is always ON, while port B acts as the input port. The truth table shows the realization of the NOT gate.

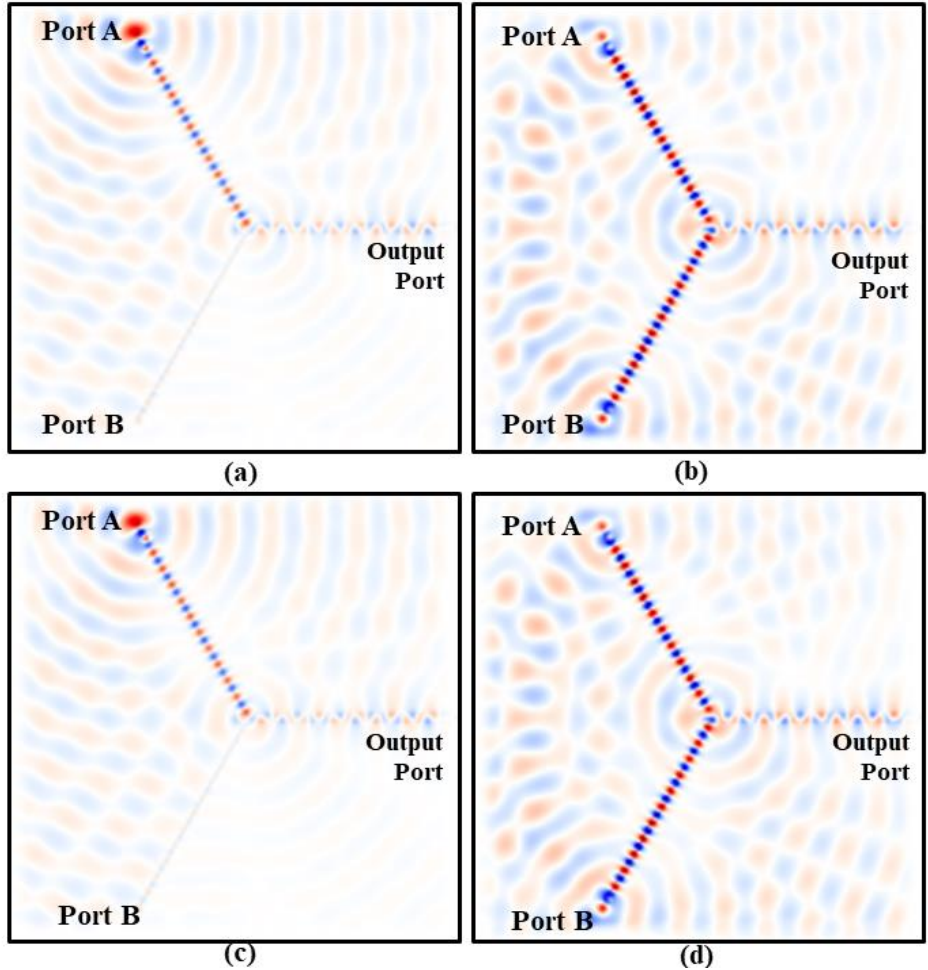

**Fig. 3:** The intensity distributions of the NOT logic gate in the TE mode of propagation (a) A=1, B=0, (b) A=1, B=1; and the TM mode of propagation (c) A=1, B=0, (d) A=1, B=1

The propagation time delay for the TE mode of the NOT gate has been determined as 8.34 femtoseconds and for the TM mode as 7.00 femtoseconds.

## Conclusions

This paper demonstrates the design of all-optical logic gates based on the superposition of light using SSWs. All the reported logic gate designs have been tested for a fan-in of 2 and a fan-out of 1. The propagation delay of electronic logic gates is of the order of nanoseconds [42, 43]; for PhC logic gates, it is of the order of a few picoseconds; while our simulations clearly establish a delay time of the order of femtoseconds. We can distinctly see that SSW-based logic gate structures demonstrate fast operation along with a simplified design as compared to the currently existing all-optical logic gates. The parameters obtained for determining the efficiency of the gates clearly establish that proposed all-optical logic gates can prove to be useful for the design and development of all-optical circuits.

## Statements and Declarations

**Disclosure Statement:** The authors report there are no competing interests to declare.

**Data Availability:** Data sharing is not applicable to this article as no datasets were generated or analyzed during the current study.

**Funding Statement:** No external funding was received for this research.